\documentclass[pre,onecolumn,amsmath,amssymb,superscriptaddress]{revtex4-2}
\usepackage{amsmath}
\usepackage{hyperref}
\usepackage{graphicx}
\usepackage{amsfonts}
\usepackage{amsthm}
\usepackage{cases}
\usepackage{bm}
\usepackage{subcaption}
\usepackage{color}
\definecolor{Blue}{rgb}{0.00, 0.00, 1.00}
\definecolor{Red}{rgb}{1.00, 0.00, 0.00}
\definecolor{Green}{rgb}{0.00, 0.70, 0.00}

\hypersetup{colorlinks=true, linkcolor=blue, citecolor=blue, filecolor=magenta, urlcolor=cyan}
\begin{document}
\title{Brownian motion near a soft surface}
\author{Yilin Ye}
\affiliation{Univ. Bordeaux, CNRS, LOMA, UMR 5798, F-33400 Talence, France.}
\affiliation{Laboratoire de Physique de la Mati\`{e}re Condens\'{e}e (UMR 7643), CNRS, Ecole Polytechnique, Institut Polytechnique de Paris, 91120 Palaiseau, France.}
\author{Yacine Amarouchene}
\email{yacine.amarouchene@u-bordeaux.fr}
\affiliation{Univ. Bordeaux, CNRS, LOMA, UMR 5798, F-33400 Talence, France.}
\author{Rapha\"el Sarfati}
\affiliation{Department of Civil and Environmental Engineering, Cornell University, Ithaca, NY, USA.}
\author{David S. Dean}
\email{david.dean@u-bordeaux.fr}
\affiliation{Univ. Bordeaux, CNRS, LOMA, UMR 5798, F-33400 Talence, France.}
\author{Thomas Salez}
\email{thomas.salez@cnrs.fr}
\affiliation{Univ. Bordeaux, CNRS, LOMA, UMR 5798, F-33400 Talence, France.}
\begin{abstract}
Brownian motion near soft surfaces is a situation widely encountered in nanoscale and biological physics. However, a complete theoretical description is lacking to date. Here, we theoretically investigate the dynamics of a two-dimensional colloid in an arbitrary external potential and near a soft surface. The latter is minimally modelled by a Winkler's foundation, and we restrict the study to the colloidal motion in the direction perpendicular to the surface. We start from deterministic hydrodynamic considerations, by invoking the already-established leading-order soft-lubrication forces acting on the particle. Importantly, a negative softness-induced and position-dependent added mass is identified. We then incorporate thermal fluctuations in the description. In particular, an effective Hamiltonian formulation is introduced and a temperature-dependent generalized potential is constructed in order to ensure equilibrium properties for the colloidal position. From these considerations and the Fokker-Planck equation, we then derive the relevant Langevin equation, which self-consistently allows to recover the deterministic equation of motion at zero temperature. Interestingly, besides an expected multiplicative-noise feature, the noise correlator appears to be modified by the surface softness. Moreover, a softness-induced temperature-dependent spurious drift term has to be incorporated within the Ito prescription. Finally, using numerical simulations with various initial conditions and parameter values, we statistically analyze the trajectories of the particle when placed within a harmonic trap and in presence of the soft surface. This allows us to: (i) quantify further the influence of surface softness, through the added mass, which enhances the velocity fluctuations; and (ii) show that intermediate-time diffusion is unaffected by softness, within the assumptions of the model.

\end{abstract}
\maketitle

\section{Introduction}
Since its first  observation and subsequent theoretical explanation~\cite{Brown1, Einstein1905, Perrin1910, Langevin1908}, Brownian motion has increasingly become a central paradigm in the field of statistical physics~\cite{dissipation, Stokes-Ein, Green, Kubo}. Adding further complexity, as inspired from microfluidic, biological and nanoscale physics, the transport properties of colloids is expected to strongly dependent on their surroundings~\cite{Mcgraw2025}. An illustration of this statement is the pioneering experimental study of colloids confined between two flat rigid walls~\cite{Faucheux1994}. In this case, the mobility is hindered as a consequence of the no-slip boundary condition~\cite{Faxen1923,Brenner1961}, which leads to spatially-inhomogeneous and anisotropic diffusion, as well as non-Gaussian displacements along the walls. There have been several experimental and numerical developments~\cite{Dufresne2007, Eral2010, Matse2017, Boniello2018,PRR2021,alexandre2023non, Millan2023} to test these ideas and the associated theoretical predictions. The hydrodynamic coupling between two Brownian particles is also modified by the nearby presence of a rigid wall~\cite{Dufresne2000}. Similarly, the velocity auto-correlation function of a single particle is modified by the presence of a rigid wall~\cite{Felderhof2005,Mo2015}. Interestingly, the modification of the diffusive properties of confined colloids can also be used to measure slip lengths at solid-liquid interfaces~\cite{Bocquet2006}. 

However, real and biologically-relevant surfaces are often soft and complex. A natural and important question thus arises in this context: what is the influence of surface softness on nearby Brownian motion? A few studies along this line of thought have been carried out for fluid interfaces~\cite{Bickel2006,Wang2009,Villa2023}, fluctuating surfaces~\cite{Marbach2018, Sarfati2021}, biological membranes~\cite{Daddi2016}, lipid bilayers~\cite{Sheikh2023}, viscoelastic walls~\cite{Diamant2017}, or a harmonic spring~\cite{lacherez2025enhanced}. Nevertheless, in most cases, a weak-coupling regime with a point-like tracer particle is considered. In contrast, the large-coupling regime where the gap between the particle and the boundary is small compared to the particle size and where the motion-induced lubrication pressure is sufficient to deform the boundary and in turn rectify the flow around the particle, has been scarcely addressed. A recent experimental work on soft colloids near a wall~\cite{fares2024observation} indicated the possible emergence of a novel force induced by the intricate coupling between thermal fluctuations, confined viscous flow and boundary elasticity, with potentially-important implications for biophysics. However, a proper statistical-physics description of such a situation is still needed.

Interestingly, the deterministic part of the problem has already been addressed in recent years. In particular, an emergent ElastoHydroDynamic (EHD) lift force at zero Reynolds number was theoretically predicted for elastic bodies moving nearby in a viscous fluid~\cite{EPL1993}. This intriguing effect and other near-contact EHD couplings have been further explored through soft-lubrication theory~\cite{beaucourt2004optimal, skotheim2004soft, skotheim2005soft,weekley2006transient, urzay2007elastohydrodynamic, trouilloud2008soft, chan2009dynamic, urzay2010asymptotic, snoeijer2013similarity, JFM2015, pandey2016lubrication, daddi2017mobility, daddi2018reciprocal, essink2021regimes, kargar2021lift, JFM2022, hu2023effect,bertin2024similarity, jha2024capillary,bha24,Dai2025}, and experiments~\cite{lequeux1992shear, vakarelski2010dynamic, leroy2012hydrodynamic, villey2013effect, kaveh2014hydrodynamic,bouchet2015experimental, wang2015out, saintyves2016self, guan2017noncontact, rallabandi2018membrane, davies2018elastohydrodynamic, vialar2019compliant,zhang2020direct,zhang2025direct}, indicating a potential relevance for biological and nanoscale systems, through, \textit{e.g.}, cartilaginous-joint lubrication, particle filtering, drug delivery, and contactless rheology. Although there are now several reviews on this area of research~\cite{wang2017elastic, karan2018small, bureau2022lift,Rallabandi2024}, and despite the potential impact on Brownian dynamics of such couplings, including further thermal fluctuations represents the crucial second part of the problem -- which remains to be addressed.

In this article, we thus theoretically investigate the Brownian dynamics of a colloid near and perpendicularly to a soft surface. Specifically, we incorporate thermal fluctuations into the previously-studied deterministic case, at leading order in soft lubrication, for a linear and local elastic response. First, we identify a softness-induced added mass term. Then, we derive an effective Langevin equation and identify the modified noise correlator due to surface softness. Besides, we find a softness-induced temperature-dependent drift, similar to the spurious drift in Ito calculus, which is necessary to enforce the Gibbs-Boltzmann distribution at equilibrium. Numerical solutions are then obtained and discussed for the case where the particle is trapped in a harmonic potential in addition to moving near the soft surface.

\section{Theory}
\subsection{System and notations}
\begin{figure}[ht!]
\begin{center}
    \includegraphics[width=0.6\linewidth]{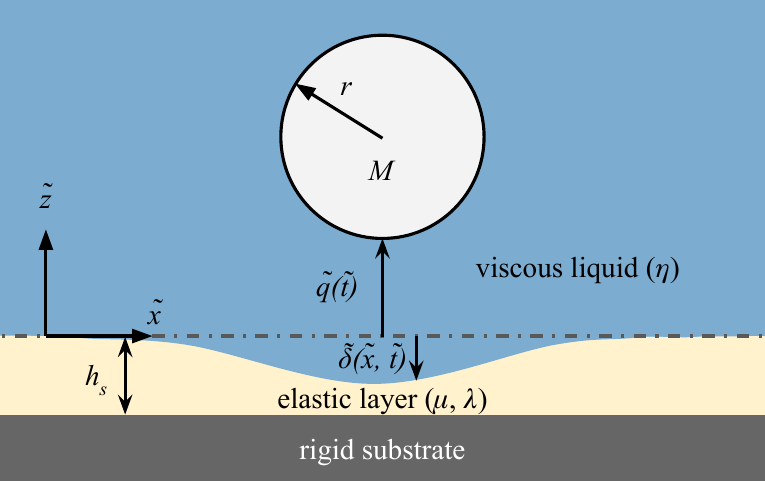}
    \caption{\textit{Schematic of the system. A two-dimensional colloidal particle (light grey) of radius $r$ and mass per unit length $M$ is free to move along the vertical direction $\tilde{z}$ and time $\tilde{t}$ under the action of thermal fluctuations, at temperature $\tilde{T}$, and an external potential $\tilde{V}(\tilde{q})$, where $\tilde{q}(\tilde{t})$ is the particle's vertical coordinate with respect to the $\tilde{z}=0$ line. The motion takes place in a Newtonian liquid (blue) of dynamic shear viscosity $\eta$, and remains close to a flat horizontal rigid substrate (dark grey) coated with a soft elastic layer (yellow) of thickness $h_{\textrm{s}}$, and Lam\'e's coefficients $\mu$ and $\lambda$. We denote by $\tilde{\delta}(\tilde{x},\tilde{t})$ the deformation field of the soft layer.
} }
\label{Fig1}
\end{center}
\end{figure}
The system at stake is shown in Fig.~\ref{Fig1}. It is inspired from~\cite{JFM2015} for the deterministic aspects that we briefly recall below, but with the key addition of thermal fluctuations. We consider a two-dimensional colloidal particle of radius $r$ and mass per unit length $M$, within a Newtonian liquid of dynamic shear viscosity $\eta$, at temperature $\tilde{T}$. We assume that the motion of the particle occurs in the vicinity of a flat horizontal rigid substrate coated with a soft elastic layer of thickness $h_{\textrm{s}}$, and Lam\'e's coefficients $\mu$ and $\lambda$. We denote by $\tilde{\delta}(\tilde{x},\tilde{t})$ the deformation field of the soft layer along the vertical coordinate $\tilde{z}$, which is a function of the horizontal coordinate $\tilde{x}$ and time $\tilde{t}$. For the sake of simplicity, we here limit ourselves to motions wherein the particle has only one degree of freedom: the gap distance $\tilde{q}(\tilde{t})$ between the particle and the undeformed boundary ($\tilde{z}=0$), along the vertical direction $\tilde{z}$. In this model, it is assumed that there is no direct energetic interaction between the soft layer and the particle, the only interaction being of hydrodynamic origin.

We further assume that $q^*=\epsilon r\ll r$, where $q^*$ is a typical scale for $\tilde{q}(\tilde{t})$ and where $\epsilon$ is the as-defined scale ratio. Due to this scale separation, we invoke the lubrication approximation, where the viscous stresses are small relative to the motion-induced hydrodynamic pressure field $\tilde{P}(\tilde{x},\tilde{t})$, which itself is independent of $\tilde{z}$ and vanishes far from the near-contact region as $\tilde{x}\rightarrow\pm\infty$. The horizontal extent $l(\tilde{t})$ of the flow-induced pressure disturbance satisfies $l(\tilde{t})\gg\tilde{q}(\tilde{t})$, and scales as $l(\tilde{t})\sim\sqrt{r\tilde{q}(\tilde{t})}\ll r$, so that, as for Hertzian contact, we can assume a parabolic shape of the liquid-gap profile $\tilde{h}(\tilde{x},\tilde{t})$, as:
\begin{equation}
\label{gapreal}
\tilde{h}(\tilde{x},\tilde{t})=\tilde{q}(\tilde{t})-\tilde{\delta}(\tilde{x},\tilde{t})+\frac{\tilde{x}^2}{2r}\ .
\end{equation}

The thin soft compressible layer is described by linear elasticity, and may also be treated via a lubrication-like theory for elastic deformations if $h_{\textrm{s}}\ll l(\tilde{t})$, leading to a Winkler-like local elastic response to the flow-induced pressure disturbance~\cite{Chandler2020}, as:
\begin{equation}
\label{locel}
\tilde{\delta}(\tilde{x},\tilde{t})=-\frac{h_{\textrm{s}}}{2\mu+\lambda}\,\tilde{P}(\tilde{x},\tilde{t})\ .
\end{equation}

To characterize the motion of the particle near the soft layer, we need to calculate the vertical drag force created by the flow-induced pressure field in the contact zone. We define the horizontal fluid velocity $\tilde{u}(\tilde{x},\tilde{z},\tilde{t})$ along $\tilde{x}$. We then non-dimensionalize the problem using the following choices: $\tilde{z}=z r\epsilon$, $\tilde{h}(\tilde{x},\tilde{t})=h(x,t) r\epsilon$, $\tilde{q}(\tilde{t})=q(t)r\epsilon$, $\tilde{x}=xr\sqrt{2\epsilon}$, $\tilde{t}=tr\sqrt{2\epsilon}/c$, $\tilde{u}(\tilde{x},\tilde{z},\tilde{t})=u(x,z,t)c$, and $\tilde{P}(\tilde{x},\tilde{t})=P(x,t)\eta c\sqrt{2}/(r\epsilon^{3/2})$, where we have introduced a velocity scale $c$ that can be, \textit{e.g.}, fixed from the mobility and the given force in a specific problem, as well as the dimensionless viscosity:
\begin{equation}
\xi=\frac{6\pi\eta r}{\epsilon Mc}\ .
\end{equation}
With these definitions, the dimensionless gap profile given by Eq.~(\ref{gapreal}) becomes:
\begin{equation}
\label{gap}
h(x,t)=q(t)+x^2+\kappa P(x,t)\ , 
\end{equation}
where the dimensionless softness is defined as:
\begin{equation}
\kappa=\frac{\sqrt{2}h_{\textrm{s}}\eta c}{r^{2}\epsilon^{5/2}(2\mu+\lambda)}\ .
\end{equation}
We assume that $\kappa \ll 1$ and restrict the solution to first perturbative order in $\kappa$. 

In the lubrication approximation, the dimensionless steady Stokes equations for incompressible viscous flow are given by:
\begin{equation}
\label{stokes}
\frac{\partial^2 u(x,z,t)}{\partial z^2}=\frac{\partial P(x,t)}{\partial x}\ ,
\end{equation}
together with no-slip boundary conditions, $u[x,z=-\kappa P(x,t),t]=0$ and $u[x,z=h(x,t)-\kappa P(x,t),t]=0$. Solving Eq.~(\ref{stokes}) with these boundary conditions, and invoking the condition of volume conservation:
\begin{equation}
\label{tfe}
\frac{\partial h(x,t)}{\partial t}+\frac{\partial}{\partial x}\int_{-\kappa P(x,t)}^{h(x,t)-\kappa P(x,t)} \textrm{d}z\ u(x,z,t)=0\ ,
\end{equation}
yields the following equation for the evolution of the gap:
\begin{equation} 
\label{eqgen3}
12\dot{q}(t)+12\kappa \frac{\partial P(x,t)}{\partial t}=\frac{\partial}{\partial x}\left[h(x,t)^3\frac{\partial P(x,t)}{\partial x}\right]\ ,
\end{equation}
where the dot corresponds to the total derivative with respect to the dimensionless time $t$.
The solution $P(x,t)$ of this equation allows us to evaluate the dimensionless vertical pressure-induced drag force per unit length $\mathcal{D}$ exerted on the particle, through: 
\begin{equation}
\label{dragp}
\mathcal{D}\simeq\int_{-\infty}^{\infty}\textrm{d}x\, P\ .
\end{equation}
Invoking the following expansion for the dimensionless pressure, $P\simeq P^{(0)}+\kappa P^{(1)}$, where $P^{(0)}|_{x\rightarrow\pm \infty} =P^{(1)}|_{x\rightarrow\pm \infty} = 0$, the soft-lubrication force per unit length can be obtained at first order in dimensionless softness $\kappa$~\cite{JFM2015}.

\subsection{Deterministic equation of motion}
At zero temperature, the dimensional equation of motion of the particle (per unit length) reads:
\begin{equation}
M\frac{\textrm{d}^2\tilde{q}}{\textrm{d}\tilde{t}^2}=\tilde{f}\ ,
\end{equation}
where $\tilde{f}$ is the total force (per unit length) exerted on the particle, which contains the previously-described soft-lubrication contribution at first order in softness, as well as any additional external force (per unit length) that derives from a potential (per unit length) $\tilde{V}(\tilde{q})$. Invoking the dimensionless variables, and defining further the dimensionless force per unit length as $f=2r\tilde{f}/(Mc^2)$, the dimensionless deterministic equation of motion reads:
\begin{equation}
\ddot{q}=f\ .
\end{equation}
By identification with Eq.~(3.7) of~\cite{JFM2015}, when the motion is solely restricted to the direction perpendicular to the wall, one has:
\begin{equation}
f=-\xi\frac{\dot{q}}{q^{3/2}}-\frac{\kappa\xi}{4}\left(21\frac{\dot{q}^2}{q^{9/2}}-\frac{15}{2}\frac{\ddot{q}}{q^{7/2}}\right)-V'(q)\ ,
\end{equation}
where $V(q)$ is the dimensionless version of the external potential per unit length $\tilde{V}(\tilde{q})$, and where the prime denotes the derivative with respect to $q$.
Therefore, the deterministic equation of motion can be rewritten as:
\begin{equation}
\label{feom}
m(q)\ddot{q}=-\xi\frac{\dot{q}}{q^{3/2}}-\frac{21\kappa\xi}{4}\frac{\dot{q}^2}{q^{9/2}}-V'(q)\ ,
\end{equation}
with an effective mass $m(q)=1+\Delta m(q)$ that includes a negative space-dependent softness-induced added mass: 
\begin{equation}
\Delta m(q)=-\frac{15\kappa\xi}{8q^{7/2}}\ .
\end{equation}
Interestingly, in an analogous fashion to the classical fluid-inertia-induced added mass~\cite{Mo2019} and its space-dependent lubrication correction~\cite{fouxon2020fluid, Zhang2023} (that we do not consider in the present study for the sake of simplicity), the viscous hydrodynamic coupling to the elastic boundary also triggers a modification of the effective mass. In itself, this is already a key feature of colloidal motion near soft surfaces. It will play a crucial role when adding thermal fluctuations in the next section. Note that, in principle, the effective mass introduced above could become negative, but that this is excluded in the small-$\kappa$ limit that we are considering.

\subsection{Langevin equation}
The central question of this study is how one can convert the deterministic equation of motion (see Eq.~(\ref{feom})) into a proper Langevin equation, or, in others words, how one can go from a zero-temperature description to a finite-temperature one. This is the object of the present section. In order to proceed, we consider an effective Hamiltonian formulation for the dimensionless position $q$ and impulsion $p$ of the particle. Given the position-dependent effective mass highlighted above, we propose to write the effective dimensionless Hamiltonian in the following form:
\begin{equation}
H(q,p) = \frac{p^2}{2m(q)} + \phi(q)\ ,
\end{equation}
where $\phi(q)$ is a generalized potential that we now aim at identifying. In the canonical ensemble, the associated equilibrium probability density function (PDF) is:
\begin{equation}
\mathcal{P}_{\textrm{eq}}(p,q) =  \frac{1}{\hat{Z}}\textrm{e}^{-\frac{H(p,q)}{T}}\ ,
\label{jpdf}
\end{equation}
where $T=2\tilde{T}k_{\textrm{B}}/(M\epsilon c^2L)$ is the dimensionless temperature, $k_{\textrm{B}}$ the Boltzmann constant, $L$ a missing length due to the two-dimensional description, and $\hat{Z}$ the normalizing partition function.
A key point in our argument is that we impose the marginal probability distribution $\bar{\mathcal{P}}_{\textrm{eq}}(q)=\int\textrm{d}p\, \mathcal{P}_{\textrm{eq}}(p,q)$ for the position $q$ to be given by the Gibbs-Boltzmann factor associated with the real external potential $V(q)$, {\em i.e.}:
\begin{equation}
\label{margreal}
\bar{\mathcal{P}}_{\textrm{eq}}(q) = \frac{1}{Z}\textrm{e}^{-\frac{V(q)}{T}}\ ,
\end{equation}
where we have renamed the partition function in order to absorb any constant factor emerging from the Gaussian integration. Evaluating the marginal probability distribution function by integrating Eq.~(\ref{jpdf}) over $p$ then gives the relation:
\begin{equation}
\phi(q) = V(q) +\frac{T}{2}\ln[m(q)]\ .
\label{peff}
\end{equation}
We thus see that, in order to ensure the correct Gibbs-Boltzmann distribution for the position, the effective Hamiltonian must include an auxiliary potential that is proportional to the temperature. 

Adding a generalized friction coefficient $\gamma(q,p)$ to the effective Hamiltonian problem above, we can write the associated Fokker-Planck equation for the time-dependent PDF $\mathcal{P}(q,p,t)$, as:
\begin{equation}
\frac{\partial \mathcal{P}(q,p,t)}{\partial t} = \frac{\partial}{\partial p} \left[T \gamma(q,p)\frac{\partial \mathcal{P}(q,p,t)}{\partial p} + \frac{\gamma(q,p)p}{m(q)}\mathcal{P}(q,p,t)+\frac{\partial H}{\partial q}\mathcal{P}(q,p,t)\right]-
\frac{\partial}{\partial q}\left[\frac{\partial H}{\partial p}\mathcal{P}(q,p,t)\right],
\end{equation}
which reduces to the Liouville equation for $\gamma(q,p)=0$. The steady-state solution of the Fokker-Planck equation is the Gibbs-Boltzmann equilibrium distribution given in Eq.~(\ref{jpdf}). In the Ito prescription of stochastic calculus~\cite{oks13}, the corresponding Langevin equation then reads: 
\begin{equation}
\dot p = -\frac{\partial H}{\partial q} -\frac{\gamma(p,q)p}{m(q)}+ T\frac{\partial \gamma(q,p)}{\partial p} +\zeta(t)\ ,\label{h1}
\end{equation}
where $\zeta(t)$ is a Gaussian white noise of zero average and a correlator given by:
\begin{equation}
\langle\zeta(t)\zeta(t')\rangle = 2T\gamma(p,q)\delta_{\textrm{D}}(t-t')\ ,
\end{equation}
where $\delta_{\textrm{D}}$ is the Dirac distribution and $\langle\rangle$ is the ensemble average. Interestingly, a spurious temperature-dependent drift term $T\partial_p \gamma(q,p)$ arises due to the Ito prescription~\cite{oks13} and the form of the effective potential (see Eq.~(\ref{peff})). 
Besides, the equation for the evolution of $q$ is:
\begin{equation}
\dot q = \frac{p}{m(q)}\ .
\label{h2}
\end{equation}

The one remaining step is now to determine the friction coefficient $\gamma(q,p)$. This is achieved by considering the zero-temperature ($T=0$) case, and identifying
Eqs.~(\ref{h1}) and~(\ref{h2}) to Eq.~(\ref{feom}). Doing this leads to the result:
\begin{equation}
\gamma(q,p) = 
  \frac{\xi}{q^{3/2}} + \frac{63\kappa\xi p}{32m(q)q^{9/2}}\ ,
\end{equation}
which at first order in $\kappa$ reduces to:
\begin{equation}
\gamma(q,p) \simeq 
  \frac{\xi}{q^{3/2}} + \frac{63\kappa\xi p}{32q^{9/2}}\ .
\end{equation}
Note that, in principle, the latter expression could become negative, but that this is once again excluded in the small-$\kappa$ limit that we are considering.

All together, one obtains the effective Langevin equation at first order in dimensionless softness $\kappa$, which reads:
\begin{equation}
m(q) \ddot q = -V'(q) -  \frac{\xi}{q^{3/2}} \dot q - \frac{21\kappa\xi}{4q^{9/2}}\dot q^2-
T\frac{21 \kappa\xi}{16 q^{9/2}}+\zeta(t) \ ,
\label{qfinal}
\end{equation} 
together with the noise correlator: 
\begin{equation}
\langle \zeta(t)\zeta(t') \rangle = 2T \left( \frac{\xi}{q^{3/2}}+\frac{63\kappa\xi}{32q^{9/2}}\dot q \right) \delta_{\textrm{D}}(t-t') \,.
\label{bruit}
\end{equation}
We stress again that, by construction, this Langevin equation self-consistently leads to the deterministic case of Eq.~(\ref{feom}) when $T=0$, and ensures the correct Gibbs-Boltzmann distribution for $q$ (see Eq.~(\ref{margreal})) at thermodynamic equilibrium.  Besides the softness-induced added mass already discussed above, and the usual multiplicative-noise feature associated with colloidal dynamics near boundaries~\cite{Faxen1923,Brenner1961,Faucheux1994,Matse2017,PRR2021,alexandre2023non}, it is interesting to note here a third striking feature associated with the hydrodynamic coupling to the soft boundary: the Langevin noise correlator appears to be modified by softness.

\section{Numerical solutions}
To explore the non-linear stochastic dynamics governed by Eqs.~(\ref{qfinal}) and (\ref{bruit}), we numerically integrate them using the Euler-Maruyama scheme. This approach allows us to properly handle the multiplicative-noise structure and the non-linear drift terms associated with the effective mass and generalized friction coefficient. For convenience, we consider an external harmonic trap, with $V(q)=k(q-q_0)^2/2$, where $k$ is the dimensionless trap stiffness and $q_0$ is the dimensionless rest position in the trap. The simulations are performed for \( \kappa \in [0, 10^{-3}] \) and \( k \in [4 \times 10^3, 4 \times 10^7] \), while keeping the other parameters fixed, with values: \( \xi = 1000 \), \( T = 1 \), and \( q_0 = 1.5 \). A time step of \(10^{-5}\) is used in all simulations. We compute averages from 100 pre-thermalized trajectories, with up to \( 2\times10^{7}\) time steps. This allows us to probe the full dynamical spectrum, from the short-time ballistic regime, through an intermediate-time diffusive regime, to long-time equilibrium. Special care is taken to ensure numerical stability and accuracy across regimes, while avoiding the effective-mass cancelling. In particular, we systematically verify that the simulations reproduce the expected equilibrium Gibbs--Boltzmann distribution for the position $q$ (see Eq.~(\ref{margreal})), thereby confirming the consistency of the model with detailed balance.
\begin{figure}[ht!]
\begin{center}
   \includegraphics[width=0.6\linewidth]{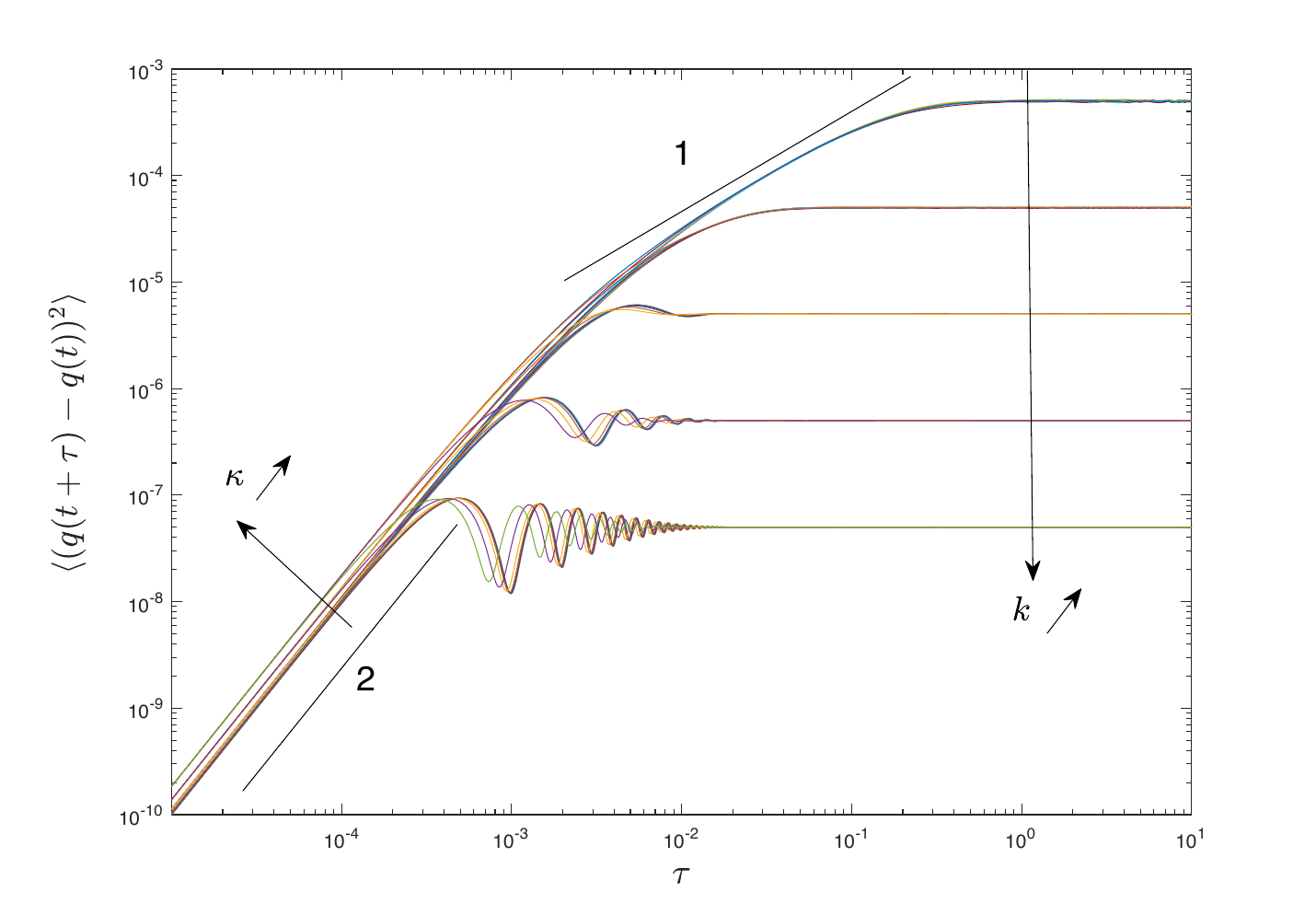}
    \caption{\textit{Dimensionless mean-square displacement (MSD) of the particle as a function of dimensionless time increment $\tau$, as obtained from statistical analysis of the numerical solutions of Eqs.~(\ref{qfinal}) and (\ref{bruit}), for dimensionless trap stiffnesses $k = 4\times10^3$, $4\times10^4$, $4\times10^5$, $4\times10^6$, $4\times10^7$, and dimensionless surface softnesses $\kappa =  0$, $1\times10^{-5}$, $2.5\times10^{-5}$, $6\times10^{-5}$, $1\times10^{-4}$, $2.5 \times10^{-4}$, $6\times10^{-4}$, $1\times10^{-3}$. The fixed parameters are the dimensionless viscosity $\xi = 1000 $, the dimensionless temperature $T = 1$, and the dimensionless rest position in the trap $q_0 = 1.5$. Straight lines indicate ballistic (2) and diffusive (1) exponents, as guides to the eye.}}
\label{Fig2}
\end{center}
\end{figure}

Figure~2 shows the vertical mean square displacement (MSD) of the particle as a function of time increment, for various values of the harmonic trap stiffness \( k \) and surface softness \( \kappa \). Each curve corresponds to a different \(\kappa\) value, spanning from the rigid case (\(\kappa = 0\)) to cases of increasingly-softer elastic layers. At small time increments, the dynamics is ballistic, characterized by a quadratic increase of the MSD versus time increment \(\tau\), consistent with inertial ballistic motion. Notably, in this regime, the MSD exhibits an increase in amplitude as \(\kappa\) increases, which is attributed to the increased velocity variance associated with the negative softness-induced added mass. At intermediate time increments, a diffusive regime is observed for the less stiff traps, characterized by a linear increase of the MSD versus time increment \(\tau\). An important result, however, is that the measured diffusion coefficient in this intermediate regime does not seem to depend on the softness $\kappa$. Hence, diffusion is not affected by soft boundaries, within the assumptions of the model, as observed in recent experiments~\cite{fares2024observation} -- despite the influence of softness on the noise correlator that we found here. The explanation probably resides in the fact that the softness-induced term in Eq.~(\ref{bruit}) is linear in velocity $\dot{q}$, and is thus averaged out at intermediate time increments. Finally, at large time increments, the MSD saturates to a plateau solely determined by the trap stiffness and temperature, confirming equilibration within the trap without noticeable effects of the surface softness.
 
\begin{figure}[ht!]
\begin{center}
    \includegraphics[width=.85\linewidth]{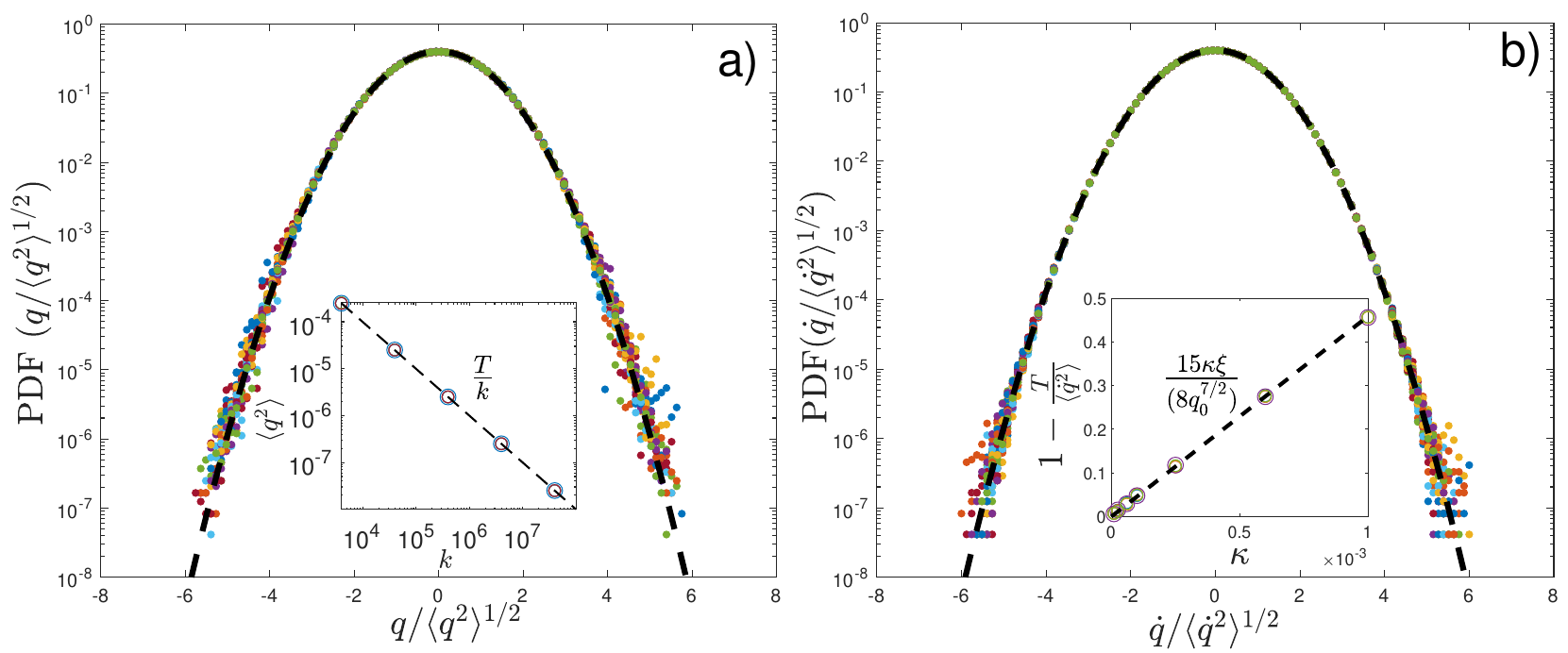}
    \caption{\textit{Equilibrium probability density functions (PDFs) of (a) the rescaled position $q/\langle q^2\rangle^{1/2}$, and (b) the rescaled velocity $\dot{q}/\langle \dot{q}^2\rangle^{1/2}$ of the particle, as obtained from statistical analysis of the numerical solutions of Eqs.~(\ref{qfinal}) and (\ref{bruit}), for dimensionless trap stiffnesses $k = 4\times10^3$, $4\times10^4$, $4\times10^5$, $4\times10^6$, $4\times10^7$, and dimensionless surface softnesses $\kappa =  0$, $1\times10^{-5}$, $2.5\times10^{-5}$, $6\times10^{-5}$, $1\times10^{-4}$, $2.5 \times10^{-4}$, $6\times10^{-4}$, $1\times10^{-3}$. The fixed parameters are the dimensionless viscosity $\xi = 1000 $, the dimensionless temperature $T = 1$, and the dimensionless rest position in the trap $q_0 = 1.5$. The dashed lines in the main plots indicate normalized Gaussian distributions. The inset in panel a) shows the variance of the position \( q \) as a function of the trap stiffness \( k \). The dashed line corresponds to the energy-equipartition prediction $\langle q^2\rangle=T/k$. The inset in panel b) shows the rescaled variance of \( \dot q \) as a function of surface softness\( \kappa \). The dashed line corresponds to the energy-equipartition prediction $\langle \dot{q}^2\rangle\simeq T/m(q_0)$ in the stiff-trap limit.}}
 \label{Fig3}
 \end{center}
\end{figure}
Figure~3 shows the equilibrium PDFs of both the rescaled particle position and velocity, for different values of the trap stiffness and surface softness. First, panel a) shows the marginal equilibrium distribution of the rescaled position. It is Gaussian, as expected for a harmonic trapping potential. The narrowing of the Gaussian width with increasing \(k\) is consistent with no adjustable parameter with energy equipartition (see inset of panel a)). Secondly, panel b) displays the distribution of the rescaled particle velocity. It is also Gaussian, as expected from the quadratic kinetic energy, but shows an interesting variance increase with $\kappa$. The latter feature is attributed to the negative softness-induced added mass. Indeed, in the stiff-trap limit where $m(q)\simeq m(q_0)$, energy equipartition captures the observed behavior with no adjustable parameter (see inset of panel b)). These results confirm the thermodynamic consistency of the numerical scheme and further illustrate the role of elastohydrodynamic interactions in modulating colloidal dynamics 

\section{Conclusion}
Through a minimal description, we theoretically and numerically investigated the dynamics of a two-dimensional colloid, placed in an arbitrary external potential and moving perpendicularly and near a soft surface within a viscous liquid. Starting from established deterministic hydrodynamic considerations, we incorporated thermal fluctuations in the description and derived the relevant Langevin equation. The latter self-consistently contains the deterministic equation of motion at zero temperature and ensures the expected thermodynamic equilibrium properties. The Langevin equations was then solved numerically for a broad range of parameter values and initial conditions. Our main findings are that: (i) a negative softness-induced added mass, reminiscent of fluid-inertial added mass, emerges due to the hydrodynamic coupling to the soft surface; (ii)  surface softness increases the velocity fluctuations through the added mass; (iii) the noise correlator is also modified by softness, but intermediate-time diffusion remains unaffected, within the assumptions of the model, which is probably due to the averaging of the fast variables; (iv) a softness-induced temperature-dependent spurious drift term appears within the Ito prescription. Future theoretical works should extend the current framework to more degrees of freedom and dimensions, incorporate more realistic non-local viscoelastic responses, including beyond-lubrication and beyond-elastic-perturbative regimes, as well as fluid inertia, and quantify the possible emergence of transient non-conservative softness-induced forces. Future experimental works should confirm the emergence of the softness-induced added mass.  

\begin{acknowledgments}
The authors thank Maxence Arutkin and Pierre Soulard for interesting discussions. They acknowledge financial support from the European Union through the European Research Council under EMetBrown (ERC-CoG-101039103) grant. Views and opinions expressed are however those of the authors only and do not necessarily reflect those of the European Union or the European Research Council. Neither the European Union nor the granting authority can be held responsible for them. The authors also acknowledge financial support from the Agence Nationale de la Recherche under EMetBrown (ANR-21-ERCC-0010-01), Softer (ANR-21-CE06-0029), and Fricolas (ANR-21-CE06-0039) grants, as well as from the Interdisciplinary and Exploratory Research Program under MISTIC grant at the University of Bordeaux, France. Finally, they thank the RRI Frontiers of Life, which received financial support from the French government in the framework of the University of Bordeaux's France 2030 program, as well as the Soft Matter Collaborative Research Unit, Frontier Research Center for Advanced Material and Life Science, Faculty of Advanced Life Science, Hokkaido University, Sapporo, Japan, and the CNRS International Research Network between France and India on ``Hydrodynamics at small scales: from soft matter to bioengineering.
\end{acknowledgments}

\section*{Data availability. } All codes used in this study to produce data are available through the online repository \url{https://github.com/EMetBrown-Lab/SoftBrownianInertia}.

\bibliography{Ye2025}
\end{document}